\documentclass[aps,prb,amsmath,twocolumn]{revtex4}

\usepackage{graphicx}
\usepackage{bm}

\begin{document}

\title{Shot noise and Coulomb blockade of Andreev reflection}

\author{Artem V. Galaktionov$^1$ and Andrei D.
Zaikin$^{2,1}$}
\affiliation{$^1$I.E. Tamm Department of
Theoretical Physics, P.N. Lebedev Physics Institute, 119991
Moscow, Russia} \affiliation{$^2$Forschungszentrum Karlsruhe,
Institut f\"ur Nanotechnologie, 76021 Karlsruhe, Germany}

\begin{abstract}
We derive low energy effective action for a short coherent
conductor between normal (N) and superconducting (S) reservoirs.
We evaluate interaction correction $\delta G$ to Andreev
conductance and demonstrate a close relation between Coulomb
effects and shot noise in NS systems. In the diffusive limit
doubling of both shot noise power and charge of the carriers
yields $|\delta G|$ four times bigger than in the normal case. Our
predictions can be directly tested in future experiments.
\end{abstract}


\maketitle

It is well known that low energy electron transport across the
interface between normal metals and superconductors (NS) is
provided by the mechanism of Andreev reflection \cite{And}. This
mechanism involves conversion of a subgap quasiparticle entering
the superconductor from the normal metal into a Cooper pair
together with simultaneous creation of a hole that goes back into
the normal metal. Each such act of electron-hole reflection
corresponds to transferring twice the electron charge $e^*=2e$
across the NS interface and results, e.g., in non-zero conductance of the
system at subgap energies \cite{BTK}.

Let us assume that two bulk metallic electrodes, one normal and
one superconducting, are connected by an arbitrary -- though
sufficiently short -- coherent conductor as it is schematically
shown in Fig. 1. This conductor is characterized by the normal
state conductance
\begin{equation}
G_N=\frac{e^2}{h}2\sum_nT_n, \label{Lf}
\end{equation}
where $T_n$ define transmissions of all conducting channels and
the factor 2 accounts for spin degeneracy. Evaluating the
conductance $G_A$ of the NS structure in Fig. 1, at
temperatures/voltages well below the superconducting gap $\Delta$
one finds \cite{BTK}
\begin{equation}
G_A=\frac{(2e)^2}{h}\sum_n{\cal T}_n, \label{BTK}
\end{equation}
where ``Andreev transmissions'' ${\cal T}_n$ are related to
$T_n$ as
\begin{equation}
{\cal T}_n=T^2_n/(2-T_n)^2. \label{Atr}
\end{equation}
Comparing Eqs. (\ref{BTK}), (\ref{Atr}) with the Landauer formula
(\ref{Lf}) one immediately observes that Andreev conductance $G_A$
formally describes ``normal'' transport of {\it spinless}
quasiparticles (hence, no extra factor 2 in front of the sum) with
charge $e^*=2e$ across some effective coherent scatterer with
channel transmissions ${\cal T}_n$ (\ref{Atr}).

Later it was realized that this formal analogy applies not only to
electron transport but also to low frequency shot noise
\cite{Khlus,dJB,MKh} and eventually to full counting statistics
(FCS) \cite{Belzig}. Consider, for instance, current fluctuations
$\delta I (t)=I(t)-I$ around its average value $I \equiv \langle
\hat I(t)\rangle$. In normal conductors at $T \to 0$ and in the
zero frequency limit the correlator for such fluctuations has the
well known form \cite{BB}
\begin{equation}
\langle |\delta I|^2\rangle =e|V|G_N\beta_N,\quad
\beta_N=\frac{\sum_nT_n(1-T_n)}{\sum_nT_n},
\label{nn}
\end{equation}
where $V$ is the average voltage across the conductor. In NS
systems Andreev reflection also leads to the current shot noise at
energies below the superconducting gap. In this case in the zero
energy/frequency limit and at $T \to 0$ one obtains \cite{dJB}
\begin{equation}
\langle |\delta I|^2\rangle =2e|V|G_A\beta_A, \quad
\beta_A=\frac{\sum_n{\cal T}_n(1-{\cal T}_n)}{\sum_n{\cal T}_n},
 \label{sn}
\end{equation}
where ${\cal T}_n$ is again defined by Eq. (\ref{Atr}). Again, a
close similarity between Eqs. (\ref{nn}) and (\ref{sn}) is
obvious: The result (\ref{sn}) just describes shot noise produced
by carriers with effective charge $e^*=2e$ in a coherent scatterer
with conductance $G_A$ and Fano factor $\beta_A$.

\begin{figure}
\includegraphics[width=6.6cm]{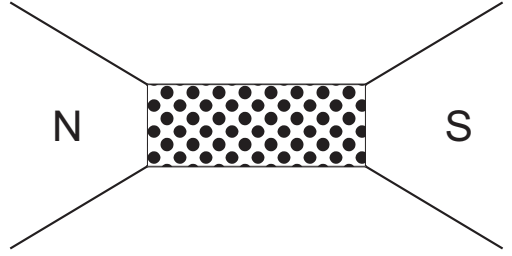}
\caption{Short coherent conductor connecting normal and
superconducting reservoirs.}
\end{figure}

In the important case of diffusive NS structures doubling of the
carrier charge also implies doubling of the shot noise \cite{dJB}.
In this case the sums over transmission channels in the above
equations can be evaluated in a straightforward manner with the
results
\begin{equation}
G_N=G_A, \quad \beta_N=\beta_A=1/3,
 \label{dFano}
\end{equation}
which yield $\langle |\delta I|^2\rangle =2e|V|G_N/3$ for NS
structures and the two times smaller result in the normal case.
This doubling of the shot noise in diffusive NS systems was indeed
observed in experiments \cite{Sanquer,Ko}.

More recently another interesting observation was reported
\cite{Bezr}. The authors of this experiment investigated short
metallic nanowires attached to bulk superconducting electrodes. In
a number of samples superconductivity inside the wire was
destroyed due to phase slippage and, hence, such samples
effectively represented hybrid normal-superconducting structures,
e.g., similar to those depicted in Fig. 1. Remarkably, the authors
\cite{Bezr} discovered that as long as the electrodes stayed
superconducting the measured I-V curves could
be well fitted by the theory of Coulomb blockade in
{\it normal} coherent conductors \cite{GZ01} provided the electron
charge $e$ was substituted by some effective charge $q_{\rm eff}$
larger than $e$ but smaller than $2e$. If, however,
superconductivity in bulk electrodes was suppressed, the I-V
curves of {\it exactly the same form} but with $q_{\rm eff} \simeq
e$ were observed. Although these observations strongly indicate
that Andreev reflection can be involved, no theoretical
explanation of the experiments \cite{Bezr} was offered until now.

Below we develop a theory describing an interplay between Coulomb
blockade and Andreev reflection. We will explicitly evaluate the
interaction correction to Andreev conductance and demonstrate its
direct correspondence to the shot noise in NS structures. Hence,
very different experiments \cite{Sanquer,Ko} and \cite{Bezr} turn
out to be closely related. Both measure the same effective charge,
i.e. $e^*=q_{\rm eff}$.

{\it Effective action}. As it is shown in Fig. 1, we will consider
big normal and superconducting reservoirs connected by a rather
short normal bridge (conductor) with arbitrary transmission
distribution $T_n$ of its conducting modes (for each $n$ the value
$T_n$ is the same for spin-up and spin-down electrons). Both phase
and energy relaxation of electrons may occur only in the
reservoirs and not inside the conductor which length is assumed to
be shorter than dephasing and inelastic relaxation lengths. In
contrast to \cite{GZ06} (where the Thouless energy
$\varepsilon_{\rm Th}$ of the normal conductor plays an important
role), here $\varepsilon_{\rm Th}$ of the bridge is irrelevant as
it is supposed to be higher than any other energy scale in our
problem. As usually, Coulomb interaction between electrons in the
conductor area is accounted for by some effective capacitance $C$.

In order to analyze electron transport in the presence of
interactions we will make use of the effective action formalism
combined with the scattering matrix technique. This approach --
very successful in the case of normal conductors
\cite{GZ01,GGZ03,KN,GZ04} -- can be conveniently generalized to
the superconducting case. In fact, the structure of the effective
action remains the same also in the latter case, one should only
replace normal propagators by $2\times2$ matrix Green functions
which account for superconductivity, as it was done, e.g., in
\cite{SZ,Z,SN}.

Following the usual procedure we express the kernel $J$ of the
evolution operator on the Keldysh contour in terms of a path
integral over the fermionic fields which can be integrated out
after the standard Hubbard-Stratonovich decoupling of the
interacting term. Then the kernel $J$ takes the form
\begin{equation}
J=\int {\cal D} \varphi_1{\cal D}\varphi_2\exp(iS[\varphi]),
\label{pathint}
\end{equation}
where $\varphi_{1,2}$ are fluctuating phases defined on the
forward and backward parts of the Keldysh contour and related to
fluctuating voltages $V_{1,2}$ across the conductor as
$\dot\varphi_{1,2}(t)=eV_{1,2}$. Here and below we set $\hbar =1$.

The effective action consists of two terms, $S[\varphi
]=S_c[\varphi ]+S_t[\varphi ]$, where
\begin{eqnarray}
iS_c[V]= \frac{C}{2e^2}\int\limits_0^t dt'
(\dot\varphi_{1}^2-\dot\varphi_{2}^2)\equiv
\frac{C}{e^2}\int\limits_0^t dt
\dot\varphi^+\dot\varphi^-\label{Sc}
\end{eqnarray}
describes charging effects and the term $S_t[V]$ accounts for
electron transfer between normal and superconducting reservoirs.
It reads \cite{SN}
\begin{equation}
S_t[\varphi]=-\frac{i}{2}\sum_n{\rm Tr} \ln \left[
1+\frac{T_n}{4}\left( \left\{ \check G_N, \check G_S
\right\}-2\right) \right],
\label{St}
\end{equation}
where $\check G_N$ and $\check G_S$ are $4\times4$ Green-Keldysh
matrices of normal and superconducting electrodes which product
implies time convolution and which anticommutator is denoted by
curly brackets. In Eq. (\ref{Sc}) we also introduced ``classical''
and ``quantum'' parts of the phase, respectively
$\varphi_+=(\varphi_1+\varphi_2)/2$ and
$\varphi_-=\varphi_1-\varphi_2$.

Without loss of generality we can set the electric potential (and,
hence, fluctuating phases) of the superconducting terminal equal
to zero. Then the Green-Keldysh matrix of this electrode can be
written in a simple form $\check G_S= \left(
\begin{smallmatrix} \hat G_R & \hat G_K \\ 0 & \hat G_A
\end{smallmatrix}\right)$ with
\begin{equation}
\hat G_{R/A}(t)=\pm \delta(t) \hat \tau_3-\theta(\pm t)\left(
\Delta \hat\tau_3 J_1(\Delta t)+i\hat\Delta J_0(\Delta t)\right)
\nonumber
\end{equation}
and $\hat G_K=\hat G_R F-F \hat G_A$, where $F(t)=-iT/\sinh [\pi T
t]$ is the Fourier transform of $1-2n(\epsilon )$ and
$n(\epsilon)=1/(1+e^{\epsilon/T})$ is the Fermi function. Here
$J_{0,1}$ are the Bessel functions, $\hat\tau_i$ are the Pauli
matrices, $\theta(t)$ is the Heaviside step function and $\hat
\Delta =i\Delta \hat\tau_2$, where $\Delta$ is chosen real.

The Green-Keldysh matrix of the normal terminal is defined as
\begin{equation}
\check G_N(t,t')=\frac{1}{2} \left( \begin{array}{cc} \hat 1& \hat
1\\\hat 1& -\hat 1 \end{array}\right) \check Q_N(t,t') \left(
\begin{array}{cc} \hat 1& \hat 1\\\hat 1& -\hat
1\end{array}\right),
\end{equation}
where
\begin{widetext}
\begin{equation}
\check Q_N (t,t')=\int \frac{d\epsilon}{2\pi}e^{-i\epsilon(t-t')}
\left(
\begin{array}{cc} e^{i\varphi_1(t)\hat\tau_3} & 0\\ 0& e^{i\varphi_2(t)\hat
\tau_3} \end{array}\right)
 \left(
\begin{array}{cc} (1-2n(\epsilon))\hat\tau_3 & 2n(\epsilon)\hat\tau_3\\
2(1-n(\epsilon))\hat\tau_3 & (2n(\epsilon)-1)\hat\tau_3
\end{array}\right) \left(
\begin{array}{cc} e^{-i\varphi_1(t')\hat\tau_3} & 0\\ 0& e^{-i\varphi_2(t')\hat
\tau_3} \end{array}\right).
\end{equation}
\end{widetext}
Substituting the above expressions for $\check G_S$ and $\check
G_N$ into Eq. (\ref{St}) we arrive at the action which fully
describes transfer of electrons between N- and S-terminals to all
orders in $T_n$.

In the limit of small channel transmissions one can expand $S_t$
in powers of $T_n$. Keeping the terms up to $\sim T_n^2$ one
recovers the contribution from Andreev reflection. At low energies
this part of the action reduces to the same form \cite{Z} as that
for normal tunnel barriers \cite{SZ} in which one substitutes $e$
by $2e$ and $G_N$ by $G_A$. Here, however, we are aiming at a more
general description which includes arbitrary transmission values
$T_n$. For this reason we will proceed differently.

Let us define the matrix $\check
X_0[\varphi_+]=1-T_n/2+(T_n/4)\left\{ \check G_N, \check G_S
\right\}|_{\varphi_-=0}$. As the action $S_t$ vanishes for
$\varphi_-(t)=0$ one has ${\rm Tr}\ln \check X_0=0$. Making use of
this property we can identically transform the action (\ref{St})
to
\begin{equation}
S_t=-\frac{i}{2}\sum_n{\rm Tr} \ln \left[ 1+\check X_0^{-1}\circ
\check X' \right], \label{xx}
\end{equation}
where $\check X'=1+(T_n/4)\left( \left\{ \check G_N, \check G_S
\right\}-2\right)-\check X_0$. At temperatures and voltages well
below the superconducting gap Andreev contribution to the action
dominates. Hence, it suffices to consider the limit of low
energies $\epsilon \ll \Delta$ and set $\check G_S\to \left(
\begin{smallmatrix} \hat\tau_2 & 0 \\ 0 & \hat\tau_2
\end{smallmatrix}\right)$.
Then we obtain
\begin{eqnarray}
&&\check X_0^{-1}(t,t')= \frac{2}{2-T_n}\label{x}\\
&&\times\left(\begin{array}{l|l}
\delta(t,t')\hat 1 &
-\frac{2T_n}{2-T_n}\sin\left[\varphi_+(t)-\varphi_+(t')\right]
F(t,t')i\hat\tau_2\\ \hline 0& \delta(t,t')\hat
1\end{array}\right),\nonumber
\end{eqnarray}
and
\begin{eqnarray}
\check X'(t,t')=\frac{T_n}{2}\delta(t,t') \left(
\begin{array}{l|l} 0& -\sin \varphi_-(t) i\hat\tau_2 \\\hline \sin
\varphi_-(t) i\hat\tau_2 & 0\end{array}\right) \nonumber
\end{eqnarray}
\begin{widetext}
\begin{eqnarray}
+T_n F(t,t')\left( \begin{array}{l|l}
-\cos\frac{\varphi_-(t)}{2}\sin\frac{\varphi_-(t')}{2}\cos(\varphi_+(t)-\varphi_+(t'))i\hat\tau_2
&  \left[
\cos\frac{\varphi_-(t)}{2}\cos\frac{\varphi_-(t')}{2}-1\right]\sin(\varphi_+(t)-\varphi_+(t'))
i\hat\tau_2 \\ \hline
\sin\frac{\varphi_-(t)}{2}\sin\frac{\varphi_-(t')}{2}\sin(\varphi_+(t)-\varphi_+(t'))i\hat\tau_2
&
\sin\frac{\varphi_-(t)}{2}\cos\frac{\varphi_-(t')}{2}\cos(\varphi_+(t)-\varphi_+(t'))
i\hat\tau_2 \end{array} \right). \label{xp}
\end{eqnarray}
\end{widetext}

Now let us assume that either dimensionless Andreev conductance
$g_A=4\sum_n{\cal T}_n$ is large, $g_A \gg 1$, or temperature is
sufficiently high (though still smaller than $\Delta$). In either
case one can describe quantum dynamics of the phase variable
$\varphi$ within the quasiclassical approximation
\cite{GZ01,GGZ03} which amounts to expanding $S_t$ in powers of
(small) ``quantum'' part of the phase $\varphi_-(t)$. Employing
Eqs. (\ref{xx})-(\ref{xp}) and expanding $S_t$ up to terms $\sim
\varphi_-^2$ we arrive at the Andreev effective action
\begin{equation}
iS_t=iS_R-S_I, \label{finalS}
\end{equation}
where
\begin{eqnarray}
iS_R&=&-\frac{ig_A}{2\pi}\;\int\limits_0^t dt'\;
\varphi^-(t')\dot\varphi^+(t'), \label{SRRR}\\
S_I&=&-\frac{g_A}{4}\int\limits_{0}^{t}dt'\int\limits_{0}^{t}dt''
\frac{T^2}{\sinh^2[\pi T(t'-t'')]} \varphi^-(t')\varphi^-(t'')
\nonumber\\
&& \times [\beta_A \cos(2\varphi^+(t')-2\varphi^+(t'')) +1-\beta_A
]. \label{SIII}
\end{eqnarray}

Eqs. (\ref{finalS})-(\ref{SIII}) represent the central result of
our work. It is remarkable that the action $S_t$ is expressed in
exactly the same form as that for normal conductors
\cite{GZ01,GGZ03} derived within the the same quasiclassical
approximation for the phase variable $\varphi (t)$. In order to
observe the correspondence between the action
\cite{GZ01,GGZ03} and that defined in Eqs.
(\ref{finalS})-(\ref{SIII}) one only needs to interchange
\begin{equation}
G_N\leftrightarrow G_A, \quad \beta_N\leftrightarrow \beta_A
\label{transform}
\end{equation}
and to account for an extra factor 2 in front of the
phase $\varphi_+$ under $\cos$ in Eq. (\ref{SIII}). This extra
factor implies doubling of the charge during Andreev reflection.

{\it Shot noise and interaction correction.} Further analysis is
formally similar to that of \cite{GZ01}. Hence, we can immediately
proceed to our final results. Let us define the average current
and the current-current correlator as
\begin{eqnarray}
&&\langle \hat I(t)\rangle =ie\int {\cal D} \varphi_{\pm}\frac{\delta}{\delta
  \varphi_-(t)}e^{iS[\varphi]}, \label{curr}\\
&&\frac12\langle \hat I\hat I\rangle_+=-e^2\int {\cal D} \varphi_{\pm}\frac{\delta^2}{\delta
  \varphi_-(t)\delta\varphi_-(t')}e^{iS[\varphi]},
\label{corr}
\end{eqnarray}
where $\langle \hat I\hat I\rangle_+=\langle \hat I(t)\hat
I(t')+\hat I(t')\hat I(t)\rangle$. In the absence of interactions
we set $\dot\varphi_{+}=eV$ and trivially recover the standard
result $I=G_AV$. For the current fluctuations $\delta I(t)$ from
Eqs. (\ref{finalS})-(\ref{corr}) analogously to \cite{GZ01} we
obtain
\begin{eqnarray}
&&\frac{\langle |\delta I|^2_\omega\rangle}{G_A}
=(1-\beta_A)\omega \coth \frac{\omega}{2T}\nonumber\\&&
+\frac{\beta_A}{2}\sum_{\pm}(\omega \pm 2eV )\coth \frac{\omega
\pm 2eV}{2T}. \label{sn1}
\end{eqnarray}
This equation fully describes current noise in NS structures at energies
well below the superconducting gap. For $eV \gg T,\omega$ Eq. (\ref{sn1})
reduces to the result \cite{dJB} (\ref{sn}) while in the diffusive regime
the correlator  (\ref{sn1}) -- together with Eqs. (\ref{dFano}) -- matches
with the semiclassical result \cite{NB}.

Let us now turn on interactions. In this case one should add the
charging term (\ref{Sc}) to the action and account for phase
fluctuations. Proceeding along the same lines as in \cite{GZ01},
for $g_A \gg 1$ or max$(T,eV) \gg E_C=e^2/2C$ we get
\begin{equation}
I=G_AV-2e\beta_A T{\rm Im}\left[ w\Psi\left(1+\frac{w}{2}
\right)-iv\Psi\left(1+\frac{iv}{2} \right) \right]. \label{iv}
\end{equation}
where $\Psi (x)$ is the digamma function, $w=g_AE_C/\pi^2T+iv$ and
$v=2eV/\pi T$. This result is plotted in Fig. 2.

\begin{figure}
\includegraphics[width=6.8cm]{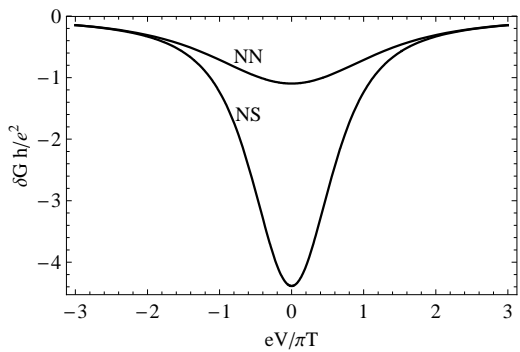}
\caption{The interaction correction $\delta G=dI/dV-G_N$ for short
diffusive conductors at $T=G_N/2\pi C$. The upper and lower curves
correspond to normal and NS structures respectively.}
\end{figure}

The last term in Eq. (\ref{iv}) is the interaction correction to
the I-V curve which scales with Andreev Fano factor $\beta_A$ in
exactly the same way as the shot noise (\ref{sn}). Thus, we arrive
at an important conclusion: {\it interaction correction to Andreev
conductance of NS structures is proportional to the shot noise
power in such structures}. This fundamental relation between
interaction effects and shot noise goes along with that
established earlier for normal conductors \cite{GZ01,LY} extending
it to superconducting systems. In both cases this relation is due
to discrete nature of the charge carriers passing through the
conductor.

Another important observation is that the interaction correction
to Andreev conductance defined in Eq. (\ref{iv}) has exactly the
same functional form as that for normal conductors, cf. Eq. (25)
in \cite{GZ01}. Furthermore, in a special case of diffusive
systems due to Eqs. (\ref{dFano}) the only difference between the
interaction corrections to the I-V curve in normal and NS systems
is the charge doubling in the latter case. As a result, the
Coulomb dip on the I-V curve of a diffusive NS system at any given
$T$ is exactly {\it 2 times narrower} than that in the normal
case. We believe that this narrowing effect was detected in normal
wires attached to superconducting electrodes \cite{Bezr}, cf. Fig.
3c in that paper \cite{FN}.

The above discussion demonstrates that seemingly different
experiments \cite{Sanquer,Ko} and \cite{Bezr} are actually closely
related: Doubling of the shot noise found in NS structures
\cite{Sanquer,Ko} corresponds to narrowing of the I-V curves
observed in \cite{Bezr}, i.e. $e^*=q_{\rm eff}$. The key reason
behind this correspondence is the relation between shot noise and
interaction correction to conductance in NS systems established
above. The absolute value of this interaction correction is
proportional to (effective charge) $\times$ (shot noise power),
i.e. doubling of the shot noise in diffusive NS structures implies
{\it 4 times bigger} interaction correction to conductance than in
the normal case, see Fig. 2. The above predictions can be verified
by independently measuring shot noise and Coulomb blockade effects
in the same NS structure, e.g., as it was already done in normal
conductors \cite{Pierre}.

In summary, we theoretically described the interplay between
Coulomb blockade and Andreev reflection and demonstrated a direct
relation between shot noise and interaction effects in NS systems.
Further extension of our theory will include the impact of
interactions on FCS.

This work was supported in part by RFBR grant 09-02-00886.


\begin{thebibliography}{99}
\bibitem{FN} The observed value of $q_{\rm eff}$ somewhat smaller than $2e$
can be attributed to additional effect of normal
quasiparticles.
\end{thebibliography}
\end{document}